\documentclass[preprint,tightenlines,aps,prd,groupedaddress,showpacs]
              {revtex4}%
\usepackage{graphicx}
\usepackage{amsmath}
\usepackage{bm}

\begin{document}

\preprint{\begin{tabular}{l}
          FERMILAB-PUB-02/274-T\\
          ANL-HEP-PR-02-094\end{tabular}}
\title{\mbox{}\\[10pt]
    Exclusive Double-Charmonium Production \\
    from $\bm{e^+ e^-}$ Annihilation into a Virtual Photon}


\author{Eric Braaten}
\affiliation{
Physics Department, Ohio State University, Columbus, Ohio 43210}
\affiliation{
Fermi National Accelerator Laboratory, P.~O.~Box 500, Batavia, Illinois 60510}

\author{Jungil Lee}
\affiliation{
Department of Physics,
Korea University,
Seoul 136-701, Korea}


\date{\today}
\begin{abstract}
We calculate the exclusive cross sections for $e^+ e^-$ annihilation
into two charmonium states through a virtual photon.
Purely electromagnetic contributions are surprisingly large, 
changing the cross sections by as much as  21\%.
The predicted cross section for $J/\psi + \eta_c$ is 
 about an order of magnitude smaller than 
a recent measurement by the BELLE Collaboration,
although part of the discrepancy can be
attributed to large relativistic corrections.
The cross sections for $S$-wave + $P$-wave, $P$-wave + $P$-wave, 
and $S$-wave + $D$-wave charmonium states are also calculated.
It may be possible to discover the $D$-wave state 
$\eta_{c2}(1D)$ at the $B$ factories through the mode
$J/\psi+\eta_{c2}$, whose cross section is predicted to be 
about a factor of 10 smaller than $J/\psi + \eta_c$. 
\end{abstract}



\maketitle


\section{Introduction
	\label{intro}}

{\it NRQCD factorization} is a systematic framework for calculating the inclusive
cross sections for producing heavy quarkonium \cite{BBL}.
The cross section for a charmonium state is expressed as the sum over 
$c \bar c$ channels of products of perturbative $c \bar c$ cross section
and nonperturbative NRQCD matrix elements.
The relative importance of the various terms in the factorization formula
is determined by the order in $\alpha_s$ of the $c \bar c$ cross section,
kinematic factors in the $c \bar c$ cross section, and 
the scaling of the NRQCD matrix element with the relative velocity $v$
of the charm quark.
Among those NRQCD matrix elements that scale with the minimal power of $v$ 
is the one associated with the color-singlet $c \bar c$ channel whose 
angular momentum quantum numbers match those of the charmonium state $H$.
The old {\it color-singlet model}  for quarkonium production~\cite{CSM} 
can be defined by keeping this channel only.  

If the charmonium is the only hadron in the initial or final state,
the color-singlet model should be accurate up to corrections 
that are higher order in $v$.  
The simplest examples of such processes are electromagnetic
annihilation decays, such as $J/\psi \to e^+ e^-$ and 
$\eta_c \to \gamma \gamma$, and exclusive electromagnetic production
processes, such as $\gamma \gamma \to \eta_c$.
Another process for which the color-singlet model should be accurate
is $e^+ e^-$ annihilation into exactly two charmonia.
There are no hadrons in the initial state, 
and the absence of additional hadrons in the final state can be 
guaranteed experimentally by the monoenergetic nature of a 2-body final state.
For many charmonia $H$, the NRQCD matrix element can be determined 
from the electromagnetic annihilation decay rate of either $H$
or of another state related to $H$ by spin symmetry.
Cross sections for double-charmonium can therefore be predicted 
 up to corrections suppressed by powers of $v^2$ 
without any unknown phenomenological factors.  

One problem with $e^+ e^-$ annihilation into exclusive double charmonium 
is that the cross sections are very small at energies large enough 
to have confidence in the
predictions of perturbative QCD.  A naive estimate of the cross section 
for $J/\psi + \eta_c$ in units of the cross section for $\mu^+ \mu^-$ is
\begin{eqnarray}
R[J/\psi + \eta_c]
\; \sim \; \alpha_s^2 \left( \frac{m_c v}{ E_{\rm beam}} \right)^6 .
\label{R-psieta:est}
\end{eqnarray}
The 2 powers of $\alpha_s$ are the fewest required to produce a 
$c \bar c + c \bar c$ final state.
There is a factor of $(m_c v)^3$ associated with the wavefunction 
at the origin for each charmonium.
These factors in the numerator are compensated by factors of the
beam energy $E_{\rm beam}$ in the denominator to get a dimensionless ratio.
As an example, consider $e^+ e^-$ annihilation with center-of-mass energy
$2 E_{\rm beam} = 10.6$ GeV.  If we set $v^2 \approx 0.3$, 
$\alpha_s \approx 0.2$, and $m_c \approx 1.4$ GeV,
we get the naive estimate $R[J/\psi + \eta_c] \approx 4\times10^{-7}$.
This should be compared to the total ratio $R[{\rm hadrons}] \approx 3.6$ 
for all hadronic final states \cite{Ammar:1997sk}.
The decay of the $J/\psi$ into the easily
detectable $e^+ e^-$ or $\mu^+ \mu^-$ modes suppresses the observable cross
section by another order of magnitude.

Fortunately, the era of high-luminosity $B$ factories has made the 
measurement of such small cross sections feasible.
The BABAR and BELLE detectors have each collected 
more than $10^7$ continuum $e^+ e^-$ annihilation events 
and more than $10^8$ events on the $\Upsilon(4S)$, 
75\% of which are continuum $e^+ e^-$ annihilation events.
The BELLE Collaboration has recently measured the cross section
for $e^+ e^- \to J/\psi + \eta_c$ \cite{Abe:2002rb}.
They also saw evidence for $J/\psi + \chi_{c0}$ and 
$J/\psi + \eta_c(2S)$ events.

In this paper, we calculate the cross sections 
for exclusive double-charmonium production via $e^+ e^-$ annihilation
into a virtual photon.  This process produces only charmonium states
with opposite charge conjugation.  The cross sections for
charmonium states with the same charge conjugation, which proceed through
$e^+ e^-$ annihilation into two virtual photons,
will be presented in subsequent papers \cite{BLB}.
We carry out the calculations in the color-singlet model
including not only the diagrams of order $\alpha^2 \alpha_s^2$
but also the purely electromagnetic diagrams of order $\alpha^4$,
which are surprisingly large.
Our result for the cross section for $J/\psi + \eta_c$
is about an order of magnitude smaller than the recent measurement 
by the BELLE Collaboration,
although part of the discrepancy can be 
attributed to large relativistic corrections. 
The cross sections for $S$-wave + $P$-wave, $P$-wave + $P$-wave, 
and $S$-wave + $D$-wave charmonium states are also calculated.
The cross section for $J/\psi+\eta_{c2}(1D)$ is predicted to be 
about a factor of 10 smaller than for $J/\psi + \eta_c$,
which may be large enough for the $D$-wave state 
$\eta_{c2}(1D)$ to be discovered at the $B$ factories. 

\section{ Color-Singlet Model Calculations
	\label{sec:CSM}}

In this section, we use the color-singlet model to calculate the
cross sections for $e^+ e^-$ annihilation 
through a virtual photon
into a double-charmonium final state $H_1+H_2$.
Charge conjugation symmetry requires one of the charmonia to be a $C=-$ state
and the other to be a $C=+$ state.
The $C=-$ states with narrow widths are the $J^{PC}=1^{--}$ states $J/\psi$
and $\psi(2S)$, the $1^{+-}$ state $h_c$, and the yet-to-be-discovered
$2^{--}$ state $\psi_2(1D)$.  
The $C=+$ states with narrow widths are the $0^{-+}$ states
$\eta_c$ and $\eta_c(2S)$, the $J^{++}$ states $\chi_{cJ}(1P)$, $J=0,1,2$,
and the yet-to-be-discovered $2^{-+}$ state $\eta_{c2}(1D)$.
We express our results in terms of the ratio $R[H_1+H_2]$
defined by
\begin{eqnarray}
R[H_1 + H_2] = 
\frac{\sigma[e^+ e^- \to H_1 + H_2]}{ \sigma[e^+ e^- \to \mu^+ \mu^-]}.
\label{R-def}
\end{eqnarray}
In the text, we give only the results for $R$ summed over helicity states.
In the Appendix, we give also the angular distribution
$dR/d\cos\theta$ for each of the helicity states of $H_1$ and $H_2$.
These results may facilitate the use of partial wave analysis to resolve the
experimental double-charmonium signal into contributions from the various
charmonium states.

\subsection{Asymptotic behavior}

 When the $e^+ e^-$ beam energy $E_{\rm beam}$ is much larger than the 
charm quark mass $m_c$, the relative sizes of the various 
double-charmonium cross sections are governed largely by the number 
of kinematic suppression factors $r^2$, where the variable $r$ is defined by
\begin{eqnarray}
r^2 &=& \frac{4 m_c^2 }{ E_{\rm beam}^2}.
\label{r-def}
\end{eqnarray}
If we set $m_c = 1.4$ GeV and $E_{\rm beam} = 5.3$ GeV,
the value of this small parameter is $r^2 = 0.28$.

\begin{figure}
\includegraphics[height=12cm,angle=-90]{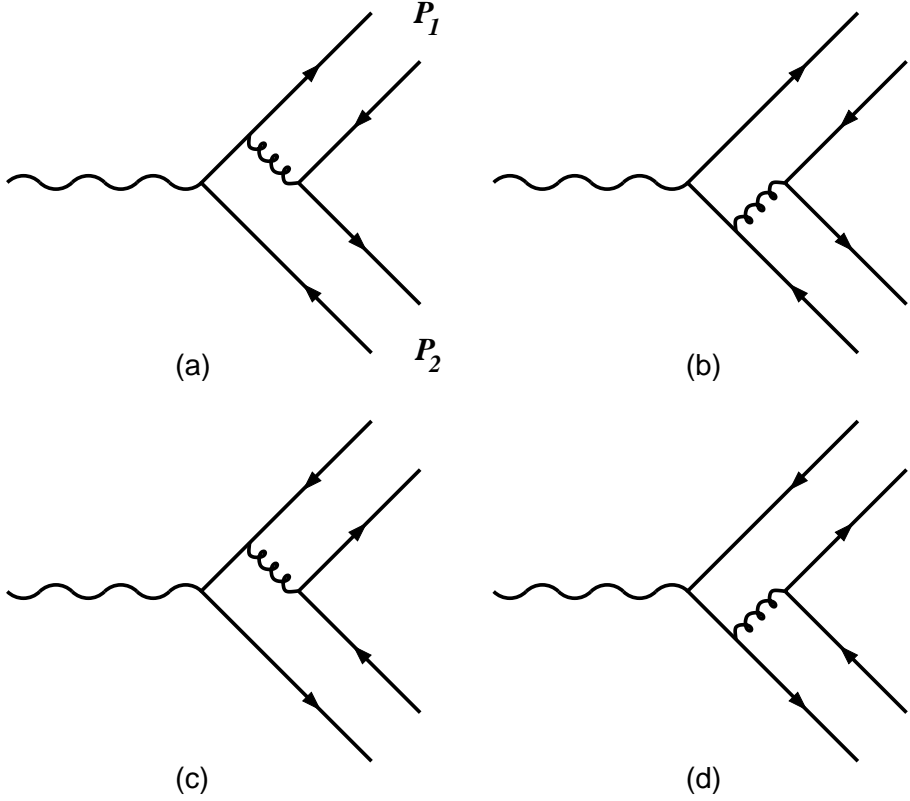}
\caption{\label{fig1}%
QCD diagrams that can contribute to the color-singlet process
$\gamma^* \to c \bar c_1 + c \bar c_1$.  }
\end{figure}

The asymptotic behavior of the ratio $R[H_1 + H_2]$ as $r \to 0$
can be determined from the {\it helicity selection rules}
for exclusive processes in perturbative QCD 
\cite{Chernyak:dj,Brodsky:1981kj}. 
For each of the $c \bar c$ pairs in the final state, 
there is a suppression factor of $r^2$ due to the large momentum transfer
required for the $c$ and $\bar c$ to emerge with small relative momentum.
Thus, at any order in $\alpha_s$, the ratio $R[H_1 + H_2]$ 
must decrease at least as fast as $r^4$ as $r \to 0$.
However it may decrease more rapidly depending on the helicity states 
of the two hadrons.
There is of course a constraint on the possible helicities
from angular momentum conservation: $|\lambda_1-\lambda_2|=0$ or 1.
The asymptotic behavior of the ratio $R[H_1(\lambda_1) + H_2(\lambda_2)]$ 
depends on the helicities $\lambda_1$ and $\lambda_2$.
The helicity selection rules imply that the slowest asymptotic decrease 
$R \sim r^4$ can occur only if the sum of the helicities of the hadrons 
is conserved.  Since there are no hadrons in the initial state,
hadron helicity conservation requires $\lambda_1 + \lambda_2 = 0$.
The only helicity state that satisfies both this constraint 
and the constraint of angular momentum conservation is 
$(\lambda_1,\lambda_2) = (0,0)$. 
For every unit of helicity by which this rule is violated, 
there is a further suppression factor of $r^2$.
The resulting estimate for the ratio $R$ at leading order in $\alpha_s$ is
\begin{eqnarray}
R_{\rm QCD}[H_1(\lambda_1) + H_2(\lambda_2)]
\; \sim \; \alpha_s^2 (v^2)^{3+L_1+L_2} (r^2)^{2+|\lambda_1+\lambda_2|}.
\label{R:lam1+lam2}
\end{eqnarray}
The factor of $v^{3+2L}$ for a charmonium state with orbital angular momentum
$L$ comes from the NRQCD factors.  At leading order of $\alpha_s$, 
there may of course be further suppression factors of $r^2$ 
that arise from the simple structure of the leading-order diagrams
for $e^+ e^- \to c \bar c_1 + c \bar c_1$ in Fig.~\ref{fig1},
but these suppression factors are unlikely to persist 
to higher orders in $\alpha_s$.

\begin{figure}
\includegraphics[height=12cm,angle=-90]{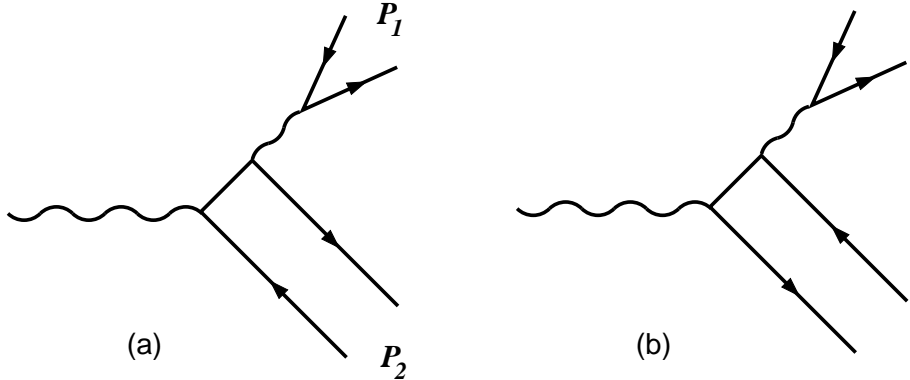}
\caption{\label{fig2}%
QED diagrams that contribute to the color-singlet process
$\gamma^* \to c \bar c_1(^3S_1) + c \bar c_1$.  }
\end{figure}

The QED diagrams for $e^+ e^- \to c \bar c_1(^3S_1) + c \bar c_1$ 
in Fig.~\ref{fig2} give contributions to 
$R[J/\psi + H_2]$ that scale in a different way with $r$.  
As $r \to 0$, the contribution to the cross section
from these diagrams factors into the cross section for $\gamma + H_2$ and the
fragmentation function for $\gamma \to J/\psi$.
This fragmentation process produces $J/\psi$ in a
$\lambda_{J/\psi} = \pm 1$ helicity state.
The hard-scattering part of the process produces only one $c \bar c$ pair 
with small relative momentum, so there is one fewer factor of $r^2$ 
relative to Eq.~(\ref{R:lam1+lam2}).  The cross section for $\gamma + H_1$ 
is still subject 
to the helicity selection rules of perturbative QCD, 
so the pure QED contribution to the ratio $R$ has the behavior
\begin{eqnarray}
R_{\rm QED}[J/\psi(\pm 1) + H_2(\lambda_2)]
\; \sim \; \alpha^2 (v^2)^{3+L_2} (r^2)^{1+|\lambda_2|}.
\label{R:psi+lam2}
\end{eqnarray}
There may also be interference terms between the QCD and QED contributions
whose scaling behavior is intermediate between
Eqs.~(\ref{R:lam1+lam2}) and (\ref{R:psi+lam2}).


\subsection{Projections onto charmonium states}

The 4 QCD diagrams for the color-singlet process 
$\gamma^* \to c \bar c_1 + c \bar c_1$ are shown in Fig.~\ref{fig1}.  
We take the upper $c \bar c$ pair in Fig.~\ref{fig1} 
to form a $C=-$ charmonium $H_1$ with momentum $P_1$ and the lower 
$c \bar c$ pair to form a $C=+$ charmonium $H_2$ with momentum $P_2$. 
There are also QED diagrams for
$\gamma^* \to c \bar c_1 + c \bar c_1$ that can be obtained from the 
QCD diagrams in Fig.~\ref{fig1} by replacing the virtual gluons 
by virtual photons, but they are suppressed by a factor of $\alpha/\alpha_s$.
However if one of the charmonia is a $1^{--}$ state like a $J/\psi$,
there are the additional QED diagrams in Fig.~\ref{fig2}.
Although they are also  suppressed by a factor of $\alpha/\alpha_s$,
they are enhanced by a kinematical factor of $1/r^2$
and therefore can be more important than one might expect.

To calculate the matrix element for $e^+ e^- \to H_1(P_1) + H_2(P_2)$,
we start from the matrix element for 
$e^+ e^- \to c(p_1) \bar c(\bar p_1) + c(p_2) \bar c(\bar p_2)$ with
the charm quarks and antiquarks on their mass shells:
$p_i^2 = \bar p_i^2 = m_c^2$.
For each of the $c \bar c$ pairs, we express the momenta in the form
\begin{subequations}
\begin{eqnarray}
p &=& \mbox{$\frac{1}{2}$} P + q,
\\
\bar p &=& \mbox{$\frac{1}{2}$} P - q,
\end{eqnarray}
\end{subequations}
where $P$ is the total momentum of the pair and
$q $ is a relative momentum that satisfies $q \cdot P = 0$.
If the $c \bar c$ pair is in a spin-singlet color-singlet state,
the matrix product of the Dirac and color spinors for the $c$ and $\bar c$
can be expressed as \cite{Bodwin:2002hg}
\begin{eqnarray}
v(\bar p) \bar u(p) =
\frac{1}{4\sqrt{2}E(E+m_c)}
\left({\overline{p} \! \! \! / \; - m_c } \right) 
\gamma_5
\left({P \! \! \! \! / + \; 2E }\right)
\left({p \! \! \!  / + \; m_c }\right)
\otimes \left( \frac{1}{\sqrt{N_c}} \, {\bf 1} \right),
\label{vubar}
\end{eqnarray}
where $E^2 = P^2/4 = m_c^2 - q^2$, $N_c = 3$, 
and the last factor involves the unit color matrix ${\bf 1}$.
If the $c \bar c$ pair is in a spin-triplet color-singlet state,
$\gamma_5$ in Eq.~(\ref{vubar}) is replaced by ${\epsilon  \! \! /}^*_S$,
where $\epsilon_S$ is a spin polarization vector 
satisfying $\epsilon_S\cdot \epsilon^*_S = -1$ 
and $P \cdot \epsilon_S = 0$.

In the spin-singlet case, the expansion of the 
matrix element in powers of $q$ has the form
\begin{eqnarray}
\mathcal{M}[c \bar c(S=0)]&=&
 \mathcal{A}
+ \mathcal{B}_{\sigma}q^\sigma
+ \mathcal{C}_{\sigma \tau} q^{\sigma} q^{\tau} + \ldots.
\label{q-expand:0}
\end{eqnarray}
The matrix elements at leading order in $v$ for the spin-singlet 
charmonium states $\eta_c$, $h_c(1P)$, and $\eta_{c2}(1D)$
can be read off from this expansion:
\begin{subequations}
\begin{eqnarray}
\mathcal{M}[\eta_c] &=&
\left(\frac{ \langle O_1 \rangle_{\eta_c} }
{ 2 N_c m_c }\right)^{1/2} \mathcal{A},
\label{M-1S0}
\\
\mathcal{M}[h_c(\lambda)] &=&
\left(\frac{ \langle O_1 \rangle_{h_c} }{ 2 N_c m_c^3 } \right)^{1/2}
\mathcal{B}_{\sigma} \epsilon^\sigma(\lambda),
\label{M-1P1}
\\
\mathcal{M}[\eta_{c2}(\lambda)]&=&
\left(\frac{ \langle O_1 \rangle_{\eta_{c2}} }{ 2 N_c m_c^5 } \right)^{1/2}
\mathcal{C}_{\sigma \tau} 
\left[ \frac{1}{2}( I^{\sigma \mu} I^{\tau \nu} + I^{\tau \mu} I^{\sigma \nu})
	-\frac{1}{3} I^{\sigma \tau} I^{\mu \nu} \right]
\epsilon_{\mu \nu}(\lambda),
\label{M-1D2}
\end{eqnarray}
\label{M-1}
\end{subequations}
where $\epsilon^\tau(\lambda)$ in Eq.~(\ref{M-1P1}) 
is a spin-1 polarization vector 
and $\epsilon_{\mu \nu}(\lambda)$ in Eq.~(\ref{M-1D2}) 
is a spin-2 polarization tensor.  
The tensor $I_{\mu \nu}$ in Eq.~(\ref{M-1D2}) is
\begin{eqnarray}
I_{\mu \nu} = -g^{\mu \nu} + \frac{P^\mu P^\nu}{4 m_c^2},
\end{eqnarray}
where $P$ is the expansion of $p + \bar p$ to leading order in $q$,
so it now satisfies $P^2 = 4 m_c^2$. 
The NRQCD matrix elements
$\langle O_1 \rangle_{\eta_c}$ and
$\langle O_1 \rangle_{h_c}$ in Eqs.~(\ref{M-1S0}) and (\ref{M-1P1})
are the vacuum-saturated analogs of the NRQCD matrix elements
$\langle O_1(^1S_0)\rangle_{\eta_c}$ and
$\langle O_1(^1P_1)\rangle_{h_c}$
for annihilation decays defined in Ref.~\cite{BBL}.
The NRQCD matrix element $\langle O_1 \rangle_{\eta_{c2}}$ in 
Eq.~(\ref{M-1D2})
is the vacuum-saturated analog of an NRQCD matrix element
$\langle O_1(^1D_2)\rangle_{\eta_{c2}}$ which in the notation 
of Ref.~\cite{BBL} is defined by
\begin{eqnarray}
\langle O_1(^1D_2)\rangle_{\eta_{c2}}
=
\langle \eta_{c2}|
\psi^\dagger
(\mbox{$-\frac{i}{2}$})^2
\mbox{$ \stackrel{\leftrightarrow}{D}{}^{(i}
        \stackrel{\leftrightarrow}{D}{}^{j)} $}
\chi
\chi^\dagger
(\mbox{$-\frac{i}{2}$})^2
\mbox{$ \stackrel{\leftrightarrow}{D}{}^{(i}
        \stackrel{\leftrightarrow}{D}{}^{j)} $}
\psi |\eta_{c2}\rangle.
\end{eqnarray}
A projection onto the $D$-wave state that is equivalent to Eq.~(\ref{M-1D2}) 
but expressed in terms of 
wavefunctions at the origin is given in Ref.~\cite{Bergstrom:1990kf}.

In the spin-triplet case, the expansion of the 
matrix element in powers of $q$ has the form
\begin{eqnarray}
\mathcal{M}[c \bar c(S=1)]&=&
\left( 
 \mathcal{A}_\rho
+\mathcal{B}_{\rho \sigma} q^\sigma
+\mathcal{C}_{\rho \sigma \tau} q^{\sigma} q^{\tau} + \ldots
\right) \epsilon_S^\rho.
\label{q-expand:1}
\end{eqnarray}
The matrix elements at leading order in $v$ for the spin-triplet 
charmonium states $J/\psi$, $\chi_{cJ}(1P)$,  $\psi_1(1D)$, 
and $\psi_2(1D)$
can be read off from this expansion:
\begin{subequations}
\begin{eqnarray}
\mathcal{M}[J/\psi(\lambda)] &=&
\left(\frac{ \langle O_1 \rangle_{J/\psi} }
{ 2 N_c m_c }\right)^{1/2}
 \mathcal{A}_\rho \epsilon^\rho(\lambda),
\label{M-3S1}
\\
\mathcal{M}[\chi_{c0}] &=&
\left(\frac{ \langle O_1 \rangle_{\chi_{c0}} }{ 2 N_c m_c^3 } \right)^{1/2}
\mathcal{B}_{\rho\sigma} \; \frac{1}{\sqrt{3}} \, I^{\rho\sigma},
\label{M-3P0}
\\
\mathcal{M}[\chi_{c1}(\lambda)] &=&
\left(\frac{ \langle O_1 \rangle_{\chi_{c1}} }{ 2 N_c m_c^3 } \right)^{1/2}
\mathcal{B}_{\rho\sigma} \;
\frac{i}{2m_c\sqrt{2}} \, \epsilon^{\rho\sigma \lambda \mu}P_{\lambda}
	\epsilon_\mu(\lambda),
\label{M-3P1}
\\
\mathcal{M}[\chi_{c2}(\lambda)] &=&
\left(\frac{ \langle O_1 \rangle_{\chi_{c2}} }{ 2 N_c m_c^3 } \right)^{1/2}
\mathcal{B}_{\rho\sigma} \;
\left[ \frac{1}{2}( I^{\rho \mu} I^{\sigma \nu} + I^{\sigma \mu} I^{\rho \nu})
	-\frac{1}{3} I^{\rho\sigma} I^{\mu \nu} \right]
\epsilon_{\mu \nu}(\lambda),
\label{M-3P2}
\\
\mathcal{M}[\psi_1(\lambda)] &=&
\left(\frac{ \langle O_1 \rangle_{\psi_2} }{ 2 N_c m_c^5 } \right)^{1/2}
\mathcal{C}_{\rho \sigma \tau}
\sqrt{\frac{3}{5}} \:
\left[ \frac{1}{2}( I^{\rho \sigma} I^{\tau \mu} + I^{\rho \tau} I^{\sigma \mu})
        -\frac{1}{3} I^{\rho\mu} I^{\sigma \tau} \right]
\epsilon_{\mu}(\lambda), 
\label{M-3D1}
\\
\mathcal{M}[\psi_2(\lambda)] &=&
\left(\frac{ \langle O_1 \rangle_{\psi_2} }{ 2 N_c m_c^5 } \right)^{1/2}
\mathcal{C}_{\rho \sigma \tau}
\frac{i}{2m_c\sqrt{6}} \:
\left[ g^{\mu \sigma}   \epsilon^{\tau \rho \lambda \nu}
     + g^{\mu \tau}   \epsilon^{\sigma \rho \lambda \nu} \right]
P_\lambda \epsilon_{\mu \nu}(\lambda).
\label{M-3D2}
\end{eqnarray}
\label{M-3}
\end{subequations}
The NRQCD matrix elements
$\langle O_1 \rangle_{J/\psi}$ and $\langle O_1 \rangle_{\chi_{cJ}}$
are the vacuum-saturated analogs of the NRQCD matrix elements
$\langle O_1(^3S_1)\rangle_{J/\psi}$ and
$\langle O_1(^3P_J)\rangle_{\chi_{cJ}}$
for annihilation decays defined in Ref.~\cite{BBL}.
The NRQCD matrix elements $\langle O_1 \rangle_{\psi_1}$ and 
$\langle O_1 \rangle_{\psi_2}$
are the vacuum-saturated analog of the NRQCD matrix elements
$\langle O_1(^3D_1)\rangle_{\psi_1}$ and 
$\langle O_1(^3D_2)\rangle_{\psi_2}$
which are defined by
\begin{subequations}
\begin{eqnarray}
\langle O_1(^3D_1)\rangle_{\psi_1}
&=&
\frac{3}{5}\,
\langle \psi_1|
\psi^\dagger
(\mbox{$-\frac{i}{2}$})^2
\sigma^i\! \stackrel{\leftrightarrow}{D}\!{}^{(i}\! 
	\stackrel{\leftrightarrow}{D}\!{}^{k)}
\chi \chi^\dagger
(\mbox{$-\frac{i}{2}$})^2
\sigma^j\! \stackrel{\leftrightarrow}{D}\!{}^{(j}\! 
	\stackrel{\leftrightarrow}{D}\!{}^{k)}
\psi|\psi_1\rangle,
\\
\langle O_1(^3D_2)\rangle_{\psi_2}
&=&
\frac{2}{3}\,
\epsilon^{iab}
\epsilon^{icd}
\langle \psi_2|
\psi^\dagger
(\mbox{$-\frac{i}{2}$})^2
\sigma^{a}\! 
\stackrel{\leftrightarrow}{D}\!{}^{(b}\!
\stackrel{\leftrightarrow}{D}\!{}^{j)}
\chi
\chi^\dagger
(\mbox{$-\frac{i}{2}$})^2
\sigma^{c}\! 
\stackrel{\leftrightarrow}{D}\!{}^{(d}\!
\stackrel{\leftrightarrow}{D}\!{}^{j)}
\psi
|\psi_2\rangle.
\end{eqnarray}
\end{subequations}

\subsection{$\bm{S}$-wave + $\bm{S}$-wave}

The matrix element for $e^+(k_2) e^-(k_1) \to J/\psi + \eta_c$
can be written as
\begin{eqnarray}
{\cal M} &=& - \frac{e_c e^2 }{ s} \,
\bar v(k_2) \gamma^\mu u(k_1) \,
\langle J/\psi+\eta_c | J_\mu(0) | \emptyset \rangle,
\end{eqnarray}
where $e_c =+\frac{2}{ 3}$ is the electric charge of the charm quark
and $s = 4 E_{\rm beam}^2$ is the square of the center-of-mass energy.
Upon simplifying the vacuum-to-$J/\psi+\eta_c$ matrix element, 
it reduces to the general form required by Lorentz covariance:
\begin{eqnarray}
\langle J/\psi(P_1,\epsilon)+\eta_c(P_2) | J_\mu(0) | \emptyset \rangle
&=& i A \, \epsilon_{\mu \nu \lambda \sigma} 
P_1^\nu P_2^\lambda \epsilon^\sigma,
\label{amp}
\end{eqnarray}
where the coefficient $A$ is 
\begin{eqnarray}
A &=& \frac{128 \pi \alpha_s }{ s^2}
\left( {\langle O_1 \rangle_{J/\psi} } 
       {\langle O_1 \rangle_{\eta_c} } \right)^{1/2}
\left( \frac{N_c^2 -1 }{ 2 N_c^2} + \frac{e_c^2 \alpha }{ N_c \alpha_s}
        + \frac{1 }{ r^2} \frac{e_c^2 \alpha }{ \alpha_s} \right).
\end{eqnarray}
After squaring the amplitude and integrating over phase space,
we obtain our final result for the ratio $R$ defined in Eq.~(\ref{R-def}):
\begin{eqnarray}
R[J/\psi + \eta_c] = 
\frac{2  \pi^2 \alpha_s^2 }{ 9} \,
X^2 (r^2-Y)^2 r^2 (1-r^2)^{3/2} \,
\frac{\langle O_1 \rangle_{J/\psi} \langle O_1 \rangle_{\eta_c} }{ m_c^6},
\label{R:psi+eta}
\end{eqnarray}
where the coefficients $X$ and $Y$ are
\begin{subequations}
\begin{eqnarray}
X &=& \frac{4 }{ 9} \left( 1 + \frac{\alpha }{ 3 \alpha_s} \right),
\\
Y &=& - \frac{\alpha }{ \alpha_s} 
     \left( 1 + \frac{\alpha }{ 3 \alpha_s} \right)^{-1}.
\end{eqnarray}
\end{subequations}
If we set $\alpha_s = 0.21$, 
their numerical values are $X=0.450$ and $Y = -0.0344$. 
Note that the ratio (\ref{R:psi+eta}) 
depends on the charm quark mass $m_c$ explicitly and also
through the variable $r$ defined in Eq.~(\ref{r-def}).
The $\alpha^2 \alpha_s^2$ term in the cross section for $e^+ e^- \to J/\psi+\eta_c$
was calculated previously by Brodsky and Ji \cite{Brodsky:1985cr}.
They presented their result in the form of a graph of $R$ versus $1/r^2$,
but they did not give an analytic expression for the cross section.

The only helicity states that contribute to Eq.~(\ref{R:psi+eta})
at this order in $\alpha_s$ are $(\lambda_1,\lambda_2) = (\pm 1,0)$,
which violate hadron helicity conservation by 1 unit.
The QCD contribution to $R$ scales like $\alpha_s^2 v^6 r^6$ 
in accord with Eq.~(\ref{R:lam1+lam2}).
There is no reason to expect the amplitude for 
the hadron-helicity-conserving state $(0,0)$ to vanish at next-to-leading 
order in $\alpha_s$, so the asymptotic behavior of the QCD contribution 
as $r \to 0$ is probably $R \sim \alpha_s^4 v^6 r^4$.

The pure QED contribution from the diagrams in Fig.~\ref{fig2}
scale like $\alpha^2 v^6 r^2$ 
in accord with Eq.~(\ref{R:psi+lam2}).  The interference term is suppressed 
only by $\alpha /(\alpha_s r^2)$, so the QED effects are larger than one 
might expect.
If we set $\sqrt{s} = 10.6$ GeV and $m_c = 1.4$ GeV, 
the electromagnetic correction increases the cross section by 29\%.

The factor of $(1-r^2)^{3/2}$ in Eq.~(\ref{R:psi+eta})
is the nonrelativistic limit of 
 $(P_{\rm CM}/E_{\rm beam})^3$,
where $P_{\rm CM}$ is the momentum of either charmonium 
in the center-of-momentum frame.  It can be expressed as
\begin{eqnarray}
\frac{ P_{\rm CM} }{ E_{\rm beam}}
= \frac{\lambda^{1/2}(s,M_{H_1}^2,M_{H_2}^2) }{ s},
\label{P_cm}
\end{eqnarray}
where $\lambda(x,y,z) = x^2 + y^2 + z^2 -2(xy + yz + zx)$.
In (\ref{R:psi+eta}), one factor of $P_{\rm CM}/E_{\rm beam}$ 
comes from the phase space for 
$H_1 + H_2$, while the other two come from the square of the 
amplitude (\ref{amp}).

\subsection{$\bm{S}$-wave + $\bm{P}$-wave}

The ratios $R$ for $J/\psi + \chi_{cJ}$ are
\begin{eqnarray}
R[J/\psi + \chi_{cJ}] =
\frac{ \pi^2 \alpha_s^2 }{ 432} \,
X^2 F_J(r,Y) r^2 (1-r^2)^{1/2} \,
\frac{\langle O_1 \rangle_{J/\psi} \langle O_1 \rangle_{\chi_{cJ}} }{ m_c^8} ,
\label{R:psi+chi}
\end{eqnarray}
where the functions $F_J(r,Y)$ are
\begin{subequations}
\begin{eqnarray}
F_0(r,Y) &=& 
	2 [ 4 Y -6 (Y+3) r^2 +7 r^4 ]^2 
	+ r^2 [ 4 + 2(Y+5) r^2 -3 r^4 ]^2,
\\
F_1(r,Y) &=&
	3 [ 8 Y - 2Y r^2 + r^4 ]^2 
	+ 3 r^2 [ 4 Y + 2(Y+2) r^2 - 3 r^4 ]^2  
\nonumber
\\
	&+& 3r^4 [ 2(3 Y + 4) -7 r^2 ]^2,
\\
F_2(r,Y) &=&
        [8 Y - 6 (Y+2) r^2 + 11 r^4  ]^2
        + 2r^2 [ 4   + 2 (Y -1) r^2 - 3 r^4 ]^2
\nonumber\\
&+&  3 r^2 [ 4 Y - 2 (Y +2) r^2 + 3 r^4 ]^2
	+ 3 r^4 [ 2(Y+2) - 5r^2 ]^2
        + 6 r^4 [ 2Y-r^2 ]^2.
\end{eqnarray}
\end{subequations}
These expressions have been expressed as sums of squares of terms that
correspond to the helicity amplitudes.
For $\chi_{c0}$ and $\chi_{c2}$, there are QCD contributions from all 
the helicity states $(\lambda_1,\lambda_2)$ 
compatible with angular momentum conservation,
 so the leading contribution scales like $\alpha_s^2 v^8 r^4$. 
For $\chi_{c1}$, there is no contribution from the 
 hadron-helicity-conserving state $(0,0)$, so
the QCD contributions are suppressed by $r^2$ 
relative to those for $\chi_{c0}$ and $\chi_{c2}$. 
The pure QED contributions to $\chi_{cJ}$ scale like $\alpha^2 v^8 r^2$
for all $J$ in accordance with Eq.~(\ref{R:psi+lam2}).
The QED contribution is suppressed by a factor 
of $\alpha^2/(\alpha_s^2 r^2)$ for $J=0$ and 2 
but only by $\alpha^2/(\alpha_s^2 r^4)$ for $J=1$.
If we set $\sqrt{s} = 10.6$ GeV and $m_c = 1.4$ GeV,
the QED corrections change the cross sections by
$+5.0$\%,  $-5.5$\%, and $+11$\% for $J=0$, 1, and 2, respectively.

The ratio $R$ for $\eta_c + h_c$ is
\begin{eqnarray}
R[\eta_c + h_c] &=& \frac{\pi^2 \alpha_s^2} {144} X^2 H(r) r^4
(1-r^2)^{1/2} \frac{\langle O_1 \rangle_{\eta_c}
                    \langle O_1 \rangle_{h_c} }{m_c^8} ~,
\label{R:etac-hc}
\end{eqnarray}
where the function $H(r)$ is
\begin{eqnarray}
H(r) &=& 2 r^2 (r^2 -2)^2 + (3 r^4 - 6 r^2 +4 )^2.
\end{eqnarray}
The dependence on $\alpha$ appears only in the overall factor $X^2$ because
the QED contribution comes from diagrams with the same topology as the
QCD diagrams in Fig.~1 of Ref.~\cite{Braaten:2002fi}.
The QED contribution from the  photon-fragmentation diagrams in Fig.~2 of
Ref.~\cite{Braaten:2002fi} vanishes because the $+$ parity of $h_c$ does  
not allow the direct coupling to a single virtual photon.
The pure QCD contribution to the result in Eq.~(\ref{R:etac-hc}) is obtained 
by substituting $X=4/9$.
This result was first calculated correctly in Ref.~\cite{Liu:2004ga}.

\subsection{$\bm{P}$-wave + $\bm{P}$-wave}

The ratios $R$ for $h_c + \chi_{cJ}$ are
\begin{eqnarray}
R[h_c + \chi_{cJ}] =
\frac{\pi^2 \alpha_s^2            }{ 108} \,
X^2 G_J(r) r^4 (1-r^2)^{3/2} \,
\frac{\langle O_1 \rangle_{h_c} \langle O_1 \rangle_{\chi_{cJ}} }{ m_c^{10}},
\label{R:h+chi}
\end{eqnarray}
where the functions $G_J(r)$ are
\begin{subequations}
\begin{eqnarray}
G_0(r) &=& 2 r^2 (6-r^2)^2,
\label{GJ0}
\\
G_1(r) &=& 24 + 78 r^2 (2-r^2)^2,
\label{GJ1}
\\
G_2(r) &=& 3 r^2 (4-5 r^2)^2 + 7 r^6.
\label{GJ2}
\label{GJ}
\end{eqnarray}
\end{subequations}
At this order in $\alpha_s$, there is no contribution to the 
cross sections for $\chi_{c0}$ and $\chi_{c2}$ from the helicity state
$(0,0)$, so the ratios $R[h_c + \chi_{cJ}]$ for $\chi_{c0}$ and $\chi_{c2}$
are suppressed relative to that for $\chi_{c1}$ by a factor
of $r^2$.  The QED contribution increases the cross section by
$2\alpha/(3\alpha_s) \approx 2 $\%.

\subsection{$\bm{S}$-wave + $\bm{D}$-wave}

The ratio $R$ for $J/\psi + \eta_{c2}$ is
\begin{eqnarray}
R[J/\psi + \eta_{c2}] &=&
\frac{4 \pi^2 \alpha_s^2           }{ 27} \,
X^2 (Y-2r^2)^2 r^2 (1-r^2)^{7/2} \,
\frac{\langle O_1 \rangle_{J/\psi} \langle O_1 \rangle_{\eta_{c2}} }{ m_c^{10}}.
\label{R:psi+Dwave}
\end{eqnarray}
At this order in $\alpha_s$ and $\alpha$, the only helicity states 
that contribute are $(\pm 1,0)$.
Thus the QCD contribution to the  ratio $R$ scales like 
 $\alpha_s^2 v^{10} r^6$  in accord with
Eq.~(\ref{R:lam1+lam2}), while the pure QED contribution scales like 
 $\alpha^2 v^{10} r^2$  in accord with Eq.~(\ref{R:psi+lam2}).
If we set $\sqrt{s} = 10.6$ GeV and $m_c = 1.4$ GeV,
the QED correction increases the cross section by about 15\%.

The ratio $R$ for $\psi_1 + \eta_c$ is
\begin{eqnarray}
R[\psi_1+\eta_c] =
\frac{\pi^2 \alpha_s^2            }{4320} \,
X^2 (26Y-21r^2+10r^4)^2 r^2 (1-r^2)^{3/2} \,
\frac{\langle O_1 \rangle_{\psi_1} \langle O_1 \rangle_{\eta_c} }%
     { m_c^{10}}.
\label{R:3D1+eta}
\end{eqnarray}
At this order in $\alpha_s$ and $\alpha$, the only helicity states 
that contribute are $(\pm 1,0)$.
Thus the QCD contribution to the  ratio $R$ scales like 
 $\alpha_s^2 v^{10} r^6$  in accord with
Eq.~(\ref{R:lam1+lam2}), while the pure QED contribution scales like 
 $\alpha^2 v^{10} r^2$  in accord with Eq.~(\ref{R:psi+lam2}).
The QED contribution increases the cross section by $41$\%.

The ratio $R$ for $\psi_2 + \eta_c$ is
\begin{eqnarray}
R[\psi_2 + \eta_c] &=&
\frac{\pi^2 \alpha_s^2            }{ 54} \,
X^2 [6+r^2(7-4r^2)^2] r^4 (1-r^2)^{3/2} \,
\frac{\langle O_1 \rangle_{\psi_2} \langle O_1 \rangle_{\eta_c} }{ m_c^{10}}.
\label{R:Dwave+eta}
\end{eqnarray}
There is a contribution from the helicity state $(0,0)$
that satisfies hadron helicity conservation, so the  ratio $R$ scales 
like $r^4$ in accord with Eq.~(\ref{R:lam1+lam2}).  
The QED contribution increases the cross section by
$2\alpha/(3\alpha_s) \approx 2$\%.

\section{ NRQCD matrix elements
	\label{sec:Pheno}}

The ratios $R$ for exclusive double-charmonium production
calculated in Section~\ref{sec:CSM} depend on 
the NRQCD matrix elements $\langle O_1 \rangle_H$.
In this section, we describe the 
phenomenological determination of these inputs.
We also give estimates based on potential models.

\subsection{Potential models}

We can obtain estimates for the NRQCD matrix elements
from the behavior of the wavefunctions near the origin in potential models.
The expressions for the NRQCD matrix elements for  
$S$-wave, $P$-wave, and $D$-wave states are
\begin{subequations}
\begin{eqnarray}
\langle O_1 \rangle_S &\approx& \frac{N_c }{ 2 \pi} |R_S(0)|^2 ,
\\
\langle O_1 \rangle_P &\approx& \frac{3 N_c }{ 2 \pi} |R_P'(0)|^2 ,
\\
\langle O_1 \rangle_D &\approx& \frac{15 N_c }{ 4 \pi} |R_D''(0)|^2.
\end{eqnarray}
\end{subequations}
The values and derivatives of the radial wavefunctions at the origin 
for four potential models are given in Ref.~\cite{Eichten:1995ch}.  
Of these four potential models, the one that is most accurate 
at short distances is the Buchm\"uller-Tye potential \cite{Buchmuller:1980su}.
The values of the NRQCD matrix elements for this potential
are given in the first column of Table~\ref{tab:O}.

\begin{table*}
\caption{\label{tab:O}The NRQCD matrix elements
        $\langle O_1 \rangle_H$ for the charmonium states $H$
        in units of (GeV)$^{2L+3}$
        where  $L=0,1,2$ for $S$-wave, $P$-wave, and $D$-wave states.
        The first column is the estimate from the Buchm\"uller-Tye
        potential model as given in Ref.~\cite{Eichten:1995ch}.
        The second and third columns are the phenomenological results
        from electromagnetic annihilation decays for $m_c = 1.4$ GeV
        at leading order (LO) and next-to-leading order (NLO) in $\alpha_s$.
        The errors are the statistical errors
        associated with the experimental inputs only.
        The {\bf bold-faced values} are used in the predictions
        for the double-charmonium cross sections.  }
\begin{ruledtabular}
\begin{tabular}{l|ccc}
    	  & \ potential model \ & \multicolumn{2}{c}{phenomenology} \\
$H$              &       &         LO        &              NLO          \\
\hline
$\eta_c$         & 0.387 & \ 0.222 $\pm$ 0.024 \ &\ 0.297  $\pm$ 0.032 \ \\
$J/\psi$         & 0.387 & 0.208 $\pm$ 0.015 & {\bf 0.335} $\pm$ 0.024   \\
\hline
$\eta_c(2S)$     & 0.253 &                   &                           \\
$\psi(2S)$       & 0.253 & 0.087 $\pm$ 0.006 & {\bf 0.139} $\pm$ 0.010   \\
\hline
$\chi_{c0}(1P)$  & 0.107 & 0.060 $\pm$ 0.015 &      0.059  $\pm$ 0.015   \\
$\chi_{c1}(1P)$, 
$h_c(1P)$        & 0.107 &                   &                           \\
$\chi_{c2}(1P)$  & 0.107 & 0.033 $\pm$ 0.006 & {\bf 0.053} $\pm$ 0.009   \\
\hline
$\psi_1(1D)$     & 0.054 & {\bf 0.095} $\pm$ 0.015 &                     \\
$\psi_2(1D)$, 
$\eta_{c2}(1D)$  & 0.054 &                   &                           \\
\end{tabular}
\end{ruledtabular}
\end{table*}

\subsection{Phenomenology}

We can obtain phenomenological values for the NRQCD matrix elements 
$\langle O_1 \rangle_{J/\psi}$ and $\langle O_1 \rangle_{\eta_c}$
from the electronic decay rate of the $J/\psi$ and
from the photonic decay rate of the $\eta_c$ \cite{BBL}.
The results for these decay rates, including the first QCD perturbative 
correction, are
\begin{subequations}
\begin{eqnarray}
\Gamma[ \eta_c \to \gamma \gamma] &=& 2 e_c^4 \pi \alpha^2 \,
\frac{\langle O_1 \rangle_{\eta_c} }{ m_c^2}
\left( 1 - \frac{20-\pi^2}{ 6} \, \frac{\alpha_s}{ \pi} \right)^2,
\label{gam-eta}
\\
\Gamma[ J/\psi \to e^+ e^-] &=& 
\frac{2 e_c^2 \pi \alpha^2}{ 3} \,
\frac{\langle O_1 \rangle_{J/\psi} }{ m_c^2}
\left( 1 - \frac{8}{3} \, \frac{\alpha_s}{ \pi} \right)^2.
\label{gam-psi}
\end{eqnarray}
\end{subequations}
We can obtain phenomenological values for the NRQCD matrix elements 
$\langle O_1 \rangle_{\chi_{c0}}$ and $\langle O_1 \rangle_{\chi_{c2}}$
from the photonic decay rates of the $\chi_{c0}$ and $\chi_{c2}$ \cite{BBL}.
The results for these decay rates, including the first QCD perturbative 
correction, are
\begin{subequations}
\begin{eqnarray}
\Gamma[ \chi_{c0} \to \gamma \gamma] &=& 6 e_c^4 \pi \alpha^2 \,
\frac{\langle O_1 \rangle_{\chi_{c0}} }{ m_c^4}
\left( 1 + \frac{3\pi^2-28 }{ 18} \, \frac{\alpha_s }{ \pi} \right)^2,
\label{gam-chi0}
\\
\Gamma[ \chi_{c2} \to \gamma \gamma] &=& \frac{8 e_c^4 \pi \alpha^2 }{ 5} \,
\frac{\langle O_1 \rangle_{\chi_{c2}} }{ m_c^4}
\left( 1 - \frac{8 }{ 3} \, \frac{\alpha_s }{ \pi} \right)^2.
\label{gam-chi2}
\end{eqnarray}
\end{subequations}
The perturbative corrections in Eqs.~(\ref{gam-eta})--(\ref{gam-chi2})
have been expressed as squares, 
because they can be calculated as corrections to the amplitudes.
We can obtain a phenomenological value for the NRQCD  matrix element
$\langle O_1 \rangle_{\psi_2(1D)}$ 
from the electronic decay rate of the $\psi_1(1D) = \psi(3770)$,
which is in the same spin-symmetry multiplet as the $\psi_2(1D)$:
\begin{eqnarray}
\Gamma[ \psi_1(1D) \to e^+ e^-] &=& 
\frac{5 e_c^2 \pi \alpha^2}{ 18} \,
\frac{\langle O_1 \rangle_{\psi_1(1D)} }{ m_c^6}.
\label{gam-psi1D}
\end{eqnarray}
The QCD perturbative correction has not been calculated.

We first determine the NRQCD matrix elements $\langle O_1 \rangle_H$
while neglecting QCD perturbative corrections.
The experimental inputs are the electronic widths of the $J/\psi$, 
$\psi(2S)$, and $\psi_1(1D)$,
the photonic width of the $\eta_c$,
and the widths and photonic branching fractions of the 
$\chi_{c0}$ and $\chi_{c2}$ \cite{PDG}.
The only other input required is the charm quark mass $m_c$.
The values of $\langle O_1 \rangle_H$ corresponding to $m_c = 1.4$ GeV
are given in the column of Table~\ref{tab:O} labelled LO. 
The error bars are the statistical errors 
associated with the experimental inputs only.
To obtain $\langle O_1 \rangle_H$ for other values of $m_c$, 
we need to multiply the values in the Table by 
$(m_c/1.4 \; {\rm GeV})^{2+2L}$. 
 
We next determine the NRQCD matrix elements $\langle O_1 \rangle_H$
including the effects of QCD perturbative corrections.
We choose the QCD coupling constant
to be $\alpha_s = 0.25$ corresponding to a renormalization scale of $2 m_c$.
The resulting values of $\langle O_1 \rangle_H$ for $m_c = 1.4$ GeV
are given in the column of Table~\ref{tab:O} labelled NLO. 
The error bars are the statistical errors 
associated with the experimental inputs only.
To obtain $\langle O_1 \rangle_H$ for other values of $m_c$, 
we need to multiply the values in the Table by 
$(m_c/1.4 \; {\rm GeV})^{2+2L}$. 
Taking into account QCD perturbative corrections 
 changes $\langle O_1 \rangle_H$
by a factor that ranges from 0.99 for $\chi_{c0}$ to 
1.61 for $J/\psi$ and $\chi_{c2}$. 
The NLO values are in closer agreement with the estimates from the
Buchm\"uller-Tye potential model than the LO values.

One complication in the determination of the NRQCD matrix element for
$\psi_1(1D)$ is that this state may have substantial mixings 
with the $\psi(2S)$ and also with continuum $D \bar D$ states.
If the mixing angle between $\psi_1(1D)$ and $\psi(2S)$ is $\phi$
and if mixing with continuum $D \bar D$ states is neglected,
the expressions for the electronic decay rates of the $\psi(2S)$ 
and $\psi_1(1D)$ are (\ref{gam-psi}) and (\ref{gam-psi1D}) 
with the substitutions
\begin{subequations}
\begin{eqnarray}
\langle O_1 \rangle_{\psi(2S)}
&\longrightarrow&
\left| \cos \phi \, \langle O_1 \rangle_{2S}^{1/2}
- \sin \phi \,\frac{\sqrt{15}}{6 m_c^2} 
	\langle O_1 \rangle_{1D}^{1/2} \right|^2,
\\
\langle O_1 \rangle_{\psi_1(1D)}
&\longrightarrow&
\left| \cos \phi \, \langle O_1 \rangle_{1D}^{1/2}
+ \sin \phi \,\frac{6 m_c^2}{\sqrt{15}}  
	\langle O_1 \rangle_{2S}^{1/2} \right|^2.
\end{eqnarray}
\end{subequations}
A recent estimate of this effect suggests 
a mixing angle $\phi = 12^\circ$ \cite{Rosner:2001nm}.
The resulting values of the NRQCD matrix elements are
$\langle O_1 \rangle_{2S} = 0.095$ GeV$^3$ and 
$\langle O_1 \rangle_{1D} = 0.013$ GeV$^7$.
This value of $\langle O_1 \rangle_{1D}$ 
is about a factor of 7 smaller than the value of
$\langle O_1 \rangle_{\psi_1(1D)}$ in Table~\ref{tab:O}.
Thus, if this mixing scenario is correct, 
the phenomenological estimate for  $\langle O_1 \rangle_{\psi_2(1D)}$ 
in Table~\ref{tab:O} could overestimate cross sections for $\psi_2(1D)$
by about a factor of 7.

Within each spin-symmetry multiplet, the NRQCD matrix elements
$\langle O_1 \rangle_H$ should have differences of order $v^2$
which we expect to be about 30\%.
We choose to use the most precise phenomenological value 
within each spin-symmetry multiplet for all members of that multiplet.
Specifically, we use the bold-faced values in the NLO column of 
Table~\ref{tab:O} for $S$-wave and $P$-wave states
and we use the bold-faced value in the LO column 
for the $D$-wave states $\eta_{c2}(1D)$, $\psi_1(1D)$, and $\psi_2(1D)$.

\section{ Relativistic corrections
	\label{sec:Rel}}

In this section, 
we calculate the relativistic corrections to the cross sections 
for the $S$-wave double charmonium.  
We also give a phenomenological determination of the NRQCD factors 
$\langle v^2 \rangle_H$ that appear in those 
relativistic corrections.

\subsection{NRQCD factor}
 
The leading relativistic correction to the CSM
amplitudes for a charmonium $H$ are conveniently expressed in terms
of a quantity that is denoted by $\langle v^2 \rangle_H$
in Ref.~\cite{BBL}.  It can be defined formally
as a ratio of matrix elements in NRQCD.  For example, in the case of 
$\eta_c$, it can be written as
\begin{eqnarray}
\langle v^2 \rangle_{\eta_c} = 
\frac{\langle \emptyset | \chi^\dagger 
	(-\frac{i}{2} {\bf D})^2 \psi | \eta_c \rangle}
	{m_c^2\,\langle \emptyset | \chi^\dagger \psi | \eta_c \rangle}.
\label{v2-eta}
\end{eqnarray}

The naive interpretation of $\langle v^2 \rangle_H$ is the
average value of $v^2$ for the charm quark in the charmonium $H$. 
In the Buchm\"uller-Tye potential model,
the average value of $v^2$ weighted by the probability density
is 0.23 for the $1S$ states 
$\eta_c$ and $J/\psi$ and 0.29 for the $2S$ states 
$\eta_c(2S)$ and $\psi(2S)$ \cite{Buchmuller:1980su}.
However the proper interpretation of the ratio of matrix elements in
(\ref{v2-eta}) is the average value of $v^2$ weighted by the wavefunction.
Unfortunately this quantity has power ultraviolet divergences
and requires a subtraction.  
For example, the wavefunction for the $1S$ states in potential models
can be approximated fairly accurately by a momentum space wavefunction
of the form $\psi(p) = 1/(p^2 + m_c^2 v_{1S}^2)^2$
where $v_{1S}$ is a phenomenological parameter.
The integral $\int d^3p \, p^2 \psi(p)$ has a linear ultraviolet divergence.
Minimal subtraction of this linear divergence gives the negative 
value $\langle v^2 \rangle_H = - 3 v_{1S}^2$.
Thus the extraction of estimates of $\langle v^2 \rangle_H$ 
from potential models is not straightforward.

There is a connection between the quantity $\langle v^2 \rangle_H$
and the mass of the
charmonium state $H$ that was first derived by Gremm and Kapustin 
\cite{Gremm:1997dq}.  The most convenient form of this relation 
for relativistic applications is
\begin{eqnarray}
M_H^2 = 4 m_c^2 \left( 1 + \langle v^2 \rangle_H + \ldots \right) ,
\label{GK}
\end{eqnarray}
where the corrections are of order $m_c^2 v^4$.
The mass $m_c$ that appears in this relation is the pole mass.
The pole mass suffers from renormalon ambiguities,
but those ambiguities are largely compensated by corresponding ambiguities
in the matrix elements that define  $\langle v^2 \rangle_H$
\cite{Braaten:1998au,Bodwin:1998mn}.

We can use the Gremm-Kapustin relation (\ref{GK})
to obtain a phenomenological determination of $\langle v^2 \rangle_H$
using the mass $M_H$ of the charmonium state as input:
\begin{eqnarray}
\langle v^2 \rangle_H \approx \frac{M_H^2 - 4 m_c^2}{4 m_c^2}.
\label{v2-def-eric}
\end{eqnarray}
where $m_c$ is the pole mass of the charm quark.
The masses for the charmonium states $\eta_c$, $J/\psi$, 
and $\psi(2S)$ are well-measured.
The $\eta_c(2S)$ was only recently discovered by the 
BELLE Collaboration with a mass of 
$M_{\eta_c(2S)} = 3654 \pm 6 \pm 8$ MeV~\cite{Choi:2002na}.
If we set $m_c=1.4$~GeV, the values of $\langle v^2 \rangle_H$
for the $S$-wave states are
$\langle v^2 \rangle_{\eta_c}     = 0.13$, 
$\langle v^2 \rangle_{J/\psi}     = 0.22$, 
$\langle v^2 \rangle_{\eta_c(2S)} = 0.70$, and
$\langle v^2 \rangle_{\psi(2S)}   = 0.73$.
The values of $\langle v^2 \rangle_H$ for the $2S$ states
are uncomfortably large, but those large values are
necessary to compensate for the fact that $2m_c=2.8$~GeV
is far from the mass of the $2S$ states. 

\subsection{Relativistic correction factor}

The relativistic correction to the cross section for the process
$J/\psi+\eta_c$ can be calculated by replacing the amplitude
factors (\ref{M-1S0}) and (\ref{M-3S1}) by
\begin{eqnarray}
\mathcal{M}[\eta_c] &=&
\left( \frac{ M_{\eta_c} \langle O_1 \rangle_{\eta_c} }
{ 4 N_c m_c^2 (1+\langle v^2\rangle_{\eta_c})} \right)^{1/2}
\left( \mathcal{A}
+\frac{m_c^2}{3} \, \langle v^2\rangle_{\eta_c} \,
\mathcal{C}_{\sigma\tau }\,I^{\sigma\tau }
\right),
\label{M-1S0:rel}
\\
\mathcal{M}[J/\psi(\lambda)] &=&
\left( \frac{ M_{J/\psi} \langle O_1 \rangle_{J/\psi} }
{ 4 N_c m_c^2 (1+\langle v^2\rangle_{J/\psi})} \right)^{1/2}
\left(
 \mathcal{A}_\rho
+
\frac{m_c^2}{3}\, \langle v^2\rangle_{J/\psi} \,
 \mathcal{C}_{\rho\sigma\tau}\,I^{\sigma\tau} 
\right)
\epsilon^\rho(\lambda).
\label{M-3S1:rel}
\end{eqnarray}
The prefactors take into account the relativistic normalizations of 
$c \bar c$ states and charmonium states.  Strictly speaking, the factor of  
$(1 + \langle v^2 \rangle_H)^{-1/2}$ should be expanded out 
to first order in $\langle v^2 \rangle_H$.  However if we use
phenomenological determinations of the NRQCD matrix elements  
$\langle O_1 \rangle_H$, the prefactor in (\ref{M-1S0:rel}) or
(\ref{M-3S1:rel}) cancels.  
We therefore choose not to expand the prefactors.

There are relativistic corrections to the electromagnetic annihilation 
decay rates used to determine the NRQCD matrix elements 
in Table~\ref{tab:O}.
For the decay rates of the $\eta_c$ and the $J/\psi$ 
given in Eqs.~(\ref{gam-eta}) and (\ref{gam-psi}),
the leading relativistic correction can be expressed as a multiplicative factor
\begin{eqnarray}
\left( 1 - \frac{1}{ 6} \, \langle v^2 \rangle_H \right)^2
\times \frac{M_H}{2 m_c(1 + \langle v^2 \rangle_H)} 
\times \frac{2 m_c}{M_H}.
\label{gam:rel}
\end{eqnarray}
The correction has been expressed as the product of three factors.
The first factor, which appears squared,
comes from the expansion of the amplitude in powers of the 
relative velocity of the $c \bar c$ pair. 
The second factor comes from the prefactor 
in Eq.~(\ref{M-1S0:rel}) or (\ref{M-3S1:rel}).
The last factor comes from the relativistic normalization factor
$1/(2 M_H)$ in the standard expression for the decay rate.
Note that the factors of $M_H$ cancel in Eq.~(\ref{gam:rel}).

We have calculated the relativistic corrections to the 
cross section for $J/\psi + \eta_c$.  The leading correction to the
ratio $R$ in Eq.~(\ref{R:psi+eta}) can be expressed as the 
multiplicative factor
\begin{eqnarray}
&&\Bigg( 1
+ \frac{8 Y + 3(Y+4)r^2 - 5r^4}{12(r^2-Y)} \langle v^2 \rangle_{J/\psi}
+ \frac{2 Y + (Y+14)r^2 - 5r^4}{12(r^2-Y)} \langle v^2 \rangle_{\eta_c} 
\Bigg)^2
\nonumber
\\
&&\times 
\frac{M_{J/\psi}}{2 m_c (1+\langle v^2 \rangle_{J/\psi})}
\frac{M_{\eta_c}}{2 m_c (1+\langle v^2 \rangle_{\eta_c})}
\times \Bigg[
1-\frac{r^2}{2(1-r^2)}
\left(\langle v^2 \rangle_{J/\psi}+\langle v^2 \rangle_{\eta_c}\right)
\Bigg]^{3/2}.
\label{rel}
\end{eqnarray}
The correction has been written as the product of three factors.
The first factor, which appears squared,  
comes from the expansion of the amplitude in powers of the 
relative velocity of the $c \bar c$ pair. 
The second factor comes from the prefactors in Eqs.~(\ref{M-1S0:rel}) and 
(\ref{M-3S1:rel}).
The last factor is the nonrelativistic expansion
of the term $(P_{\rm{CM}}/E_{\rm{beam}})^3$
divided by its value in the nonrelativistic limit,
where one power is from the phase space factor (\ref{P_cm})
and the other two are from the square of the amplitude
in Eq.~(\ref{amp}).

Our final result for the relativistic correction can be expressed as
a multiplicative factor obtained by dividing (\ref{rel})
by a factor of (\ref{gam:rel}) for each of the charmonium states.
We express it in the form:
\begin{eqnarray}
&&\Bigg( 1
+ \frac{8 Y + 3(Y+4)r^2 - 5r^4}{12(r^2-Y)} \langle v^2 \rangle_{J/\psi}
+ \frac{2 Y + (Y+14)r^2 - 5r^4}{12(r^2-Y)} \langle v^2 \rangle_{\eta_c} 
\Bigg)^2
\nonumber
\\
&&\times 
\left( 1 - \frac{1}{ 6} \, \langle v^2 \rangle_{J/\psi} \right)^{-2}
\left( 1 - \frac{1}{ 6} \, \langle v^2 \rangle_{\eta_c} \right)^{-2}
\times \frac{M_{J/\psi} M_{\eta_c}}{4 m_c^2}
\times \left(\frac{P_{\rm CM}/E_{\rm beam}}{(1-r^2)^{1/2}}\right)^3.
\label{rel-fac}
\end{eqnarray}

\section{ Predictions for $\bm{B}$ factories
	\label{sec:Bfactories}}

In this section, we calculate the cross sections for 
exclusive double-charmonium production 
in $e^+ e^-$ annihilation at the $B$ factories. 
We also give a careful analysis of the errors in the predictions
for $J/\psi + \eta_c$.

\subsection{ Cross sections
	\label{sec:sigma}}

The results in Section~\ref{sec:CSM} were expressed in terms of the 
ratio $R$ defined in Eq.~(\ref{R-def}). The corresponding cross sections are
\begin{eqnarray}
\sigma[ H_1+H_2] = \frac{4 \pi \alpha^2}{3 s} R[ H_1+H_2].
\end{eqnarray}
The ratios $R$ depend on a number of inputs: 
the coupling constants $\alpha_s$ and $\alpha$, the charm quark mass $m_c$,
and the NRQCD matrix elements $\langle O_1 \rangle_H$.

The value of the QCD coupling constant $\alpha_s$ 
depends on the choice of the scale $\mu$.
In the QCD diagrams of Fig.~\ref{fig1}, the invariant mass of the gluon is
$\sqrt{s}/2$.  We therefore choose the scale to be $\mu = 5.3$ GeV. 
The resulting value of the QCD coupling constant is 
$\alpha_s(\mu) = 0.21$. 

The numerical value for the pole mass $m_c$ of the charm quark is unstable
under perturbative corrections, so it must be treated with care. 
Since the expressions for the electromagnetic annihilation decay rates in
Eqs.~(\ref{gam-eta})--(\ref{gam-chi2}) 
include the perturbative correction of order
$\alpha_s$, the appropriate choice for the charm quark mass $m_c$ 
in these expressions is the pole mass with corrections 
of order $\alpha_s$ included.
It can be expressed as 
\begin{eqnarray}
m_c = \bar m_c(\bar m_c) \left(1 + \frac{4}{3} \frac{\alpha_s}{\pi} \right).
\end{eqnarray}
Taking the running mass of the charm quark to be
$\bar m_c(\bar m_c) = 1.2 \pm 0.2$~GeV, 
the NLO pole mass is $m_c = 1.4 \pm 0.2$~GeV.

Our predictions for the double-charmonium cross sections 
without relativistic corrections
are given in Table~\ref{tab:sigma}. 
The error bars are those associated with the uncertainty 
in the NLO pole mass $m_c$ only.

\setcounter{table}{1}
\begin{table*}
\caption{\label{tab:sigma}Cross sections in fb for $e^+ e^-$ annihilation into
        double-charmonium states $H_1+H_2$
        without relativistic corrections.
        The errors are only those from variations in the NLO pole mass
        $m_c = 1.4 \pm 0.2$ GeV. 
}
\begin{ruledtabular}
\begin{tabular}{l|ccccc}
$H_2$ $\backslash$ $H_1$
& $J/\psi$ &$\psi(2S)$ & $h_c(1P)$    & $\psi_1(1D)$ & $\psi_2(1D)$ \\
\hline
$\eta_c$
& 3.78 $\pm$ 1.26  & 1.57 $\pm$ 0.52 &  0.308 $\pm$ 0.017 &
 0.106 $\pm$ 0.025 & 1.04 $\pm$ 0.23 \\
$\eta_c(2S)$
& 1.57 $\pm$ 0.52 & 0.65 $\pm$ 0.22 & 0.128 $\pm$ 0.007 &
 0.044 $\pm$ 0.010 & 0.43 $\pm$ 0.09 \\
$\chi_{c0}(1P)$                                         
& 2.40 $\pm$ 1.02 & 1.00 $\pm$ 0.42 & 0.053 $\pm$ 0.019 &  & \\
$\chi_{c1}(1P)$                                         
& 0.38 $\pm$ 0.12 & 0.16 $\pm$ 0.05 & 0.258 $\pm$ 0.064 &  & \\
$\chi_{c2}(1P)$                                         
& 0.69 $\pm$ 0.13 & 0.29 $\pm$ 0.06 & 0.017 $\pm$  0.002 &  & \\
$\eta_{c2}(1D)$                                         
& 0.35 $\pm$ 0.05 & 0.14 $\pm$ 0.02  &                   &  & \\
\end{tabular}
\end{ruledtabular}
\end{table*}

Our predictions for the double-charmonium cross sections 
for the $S$-wave states including the leading relativistic correction
are obtained by multiplying the values in Table~\ref{tab:sigma}
by the factor (\ref{rel-fac}). 
 We use the values of $\langle v^2 \rangle_H$ obtained from 
Eq.~(\ref{v2-def-eric}), which follows from the Gremm-Kapustin relation. 
The resulting cross sections are given in Table~\ref{tab:sigma-rel}.
The error bars are those associated with the uncertainty 
in the NLO pole mass $m_c$ only.
The relativistic corrections increase the central values of the
cross sections by about 
2 for $J/\psi+\eta_c$, by about 5 for $J/\psi+\eta_c(2S)$, 
by about 4 for $\psi(2S)+\eta_c$, and by about 8 for $\psi(2S)+\eta_c(2S)$.
Although the total correction factor for $J/\psi+\eta_c$ 
is significantly larger than 1, it is the product 
of several modest correction factors that all go in the same direction.
The largest individual correction factor for $J/\psi+\eta_c$ is $(1.28)^2$ 
coming from the expansion of the amplitude.  
 The corresponding factors for $J/\psi+\eta_c(2S)$,
$\psi(2S)+\eta_c$, and $\psi(2S)+\eta_c(2S)$ are $(1.80)^2$, $(1.64)^2$, 
and $(2.16)^2$, respectively.
 These large correction factors indicate that 
the relativistic corrections to the cross sections involving $2S$ states are 
too large to be calculated reliably using the method we have chosen.

Note that our method for calculating the relativistic correction 
significantly increases the sensitivity to the charm quark mass.  
The errors from varying $m_c$ in Table~\ref{tab:sigma} are about 50\%
for the $S$-wave states, while the errors in Table~\ref{tab:sigma-rel}
correspond to increasing or decreasing the cross section by about a factor of 3.
The strong sensitivity to $m_c$ is another indication that our method 
for calculating the relativistic corrections is unreliable.
We will therefore take the values in Table~\ref{tab:sigma}
to be our predictions for the cross sections and use Table~\ref{tab:sigma-rel} 
as an indication of the possible size of the relativistic corrections.

\begin{table}
\caption{\label{tab:sigma-rel}
Cross sections in fb for $e^+ e^-$ annihilation into
        $S$-wave double-charmonium states $H_1+H_2$
        including relativistic corrections.
        The errors are only those from variations in the NLO pole mass
        $m_c = 1.4 \pm 0.2$ GeV.  }

\begin{ruledtabular}
\begin{tabular}{l|cc}
$H_2$ $\backslash$ $H_1$
&     $J/\psi$    &     $\psi(2S)$        \\
\hline
$\eta_c$
&~$7.4^{+10.9}_{-4.1}$ & ~$6.1^{+9.5}_{-3.4}$ \\
$\eta_c(2S)$
&~$7.6^{+11.8}_{-4.1}$ & ~$5.3^{+9.1}_{-2.9}$ \\
\end{tabular}
\end{ruledtabular}

\end{table}

\subsection{ Perturbative corrections
	\label{sec:perturbative}}

The QCD perturbative 
corrections to the electromagnetic annihilation decay rates 
used to determine the NRQCD matrix elements in Table~\ref{tab:O}
have already been taken into account.  The QCD 
perturbative corrections to the 
cross section for $J/\psi + \eta_c$ have not yet been calculated. 
However parts of the perturbative corrections are related to perturbative 
corrections
that have been calculated and we can use these to estimate the order of 
magnitude of the perturbative corrections.

Some of the perturbative corrections can be associated with the wavefunctions 
of the $\eta_c$ and $J/\psi$. 
For the QED diagrams in Fig.~\ref{fig2}, the QCD perturbative corrections 
would be very closely related to those in the electromagnetic 
annihilation decay rates (\ref{gam-eta}) and (\ref{gam-psi}).
However for the QCD diagrams in Fig.~\ref{fig1}, the QCD perturbative 
corrections associated with the wavefunction could be very different.
In the expressions for electromagnetic annihilation decay rates
in Eqs.~(\ref{gam-eta})--(\ref{gam-chi2}),  we have 4 examples of
perturbative corrections associated with wavefunctions.
The root-mean-square of the 4 coefficients of $\alpha_s$ is 0.66.
We will therefore take $(1\pm 0.66 \alpha_s)^2$ 
as our estimate for the perturbative correction 
associated with each charmonium wavefunction.

There are other perturbative corrections that can be associated 
with the electromagnetic charm current $\bar c \gamma^\mu c$.
As an estimate for the 
magnitude of these corrections, we can use the perturbative correction
to the inclusive charm cross section. The corresponding ratio  
$R$ is \cite{Chetyrkin:1994pi}
\begin{eqnarray}
R[c \bar c + X]
= 3 e_c^2 (1+r^2/8) (1-r^2/4)^{1/2}
\left[ 1 + 
\left( 1 + \frac{3r^2}{4}  - \frac{r^4}{16}  + \ldots \right) 
\frac{\alpha_s}{\pi} \right].
\end{eqnarray}
If we set $m_c = 1.4$ GeV and $E_{\rm beam} = 5.3$ GEV, 
then $r^2 = 0.28$ and the perturbative correction gives 
a multiplicative factor $(1 + 0.19 \alpha_s)^2$.

There are also perturbative corrections that can be associated 
with the QCD coupling constants.  We can estimate the size of these 
perturbative
corrections by varying the scale $\mu$ up or down by a factor of 2. 
The factor $\alpha_s^2$ in the cross section changes by a
multiplicative factor of $(1\pm 0.92 \alpha_s)^2$. 

To obtain an estimate of the errors in the double charmonium cross section
associated with perturbative corrections, we will add in quadrature
the coefficients of $\alpha_s$ in the perturbative correction factors 
associated with each wavefunction, the charm current, and the factors of
$\alpha_s$.  The resulting correction is 
a multiplicative factor $(1 \pm 1.3 \alpha_s)^2$.
If we set $\alpha_s = 0.21$, this factor ranges from about 0.53 to about 1.62.
Thus we should not be surprised if the QCD perturbative corrections
changed the predictions by 60\%.

\subsection{ Color-octet contributions
	\label{sec:color-octet}}

According to the NRQCD factorization formalism \cite{BBL},
the inclusive double-charmonium cross sections are obtained by replacing
the decay NRQCD matrix elements in the color-singlet model terms
by production NRQCD matrix elements and by adding additional terms 
involving color-octet matrix elements.
The color-octet contributions at this order in $\alpha_s$ 
can be obtained from the results in Section~\ref{sec:CSM}
by replacing the NRQCD matrix elements
by appropriate color-octet matrix elements and by replacing the constants 
$X$ and $Y$ by the constants $X_8 = 1/\sqrt{72}$ and $Y_8 = -3$. 

As an illustration, we consider the inclusive production of $J/\psi + \eta_c$.  
The leading color-octet contribution for $J/\psi + \eta_c + X$
can be obtained from the color-singlet model result for 
$\chi_{cJ} + h_c$ in Eq.~(\ref{R:h+chi}) by substituting
$\langle O_1(^3P_J) \rangle_{\chi_{cJ}} 
	\to \langle O_8^{J/\psi}(^3P_J) \rangle/(2J+1)$,
$\langle O_1(^1P_1) \rangle_{h_c} \to \langle O_8^{\eta_c}(^1P_1) \rangle$,
$X \to X_8$ and $Y \to Y_8$.
The ratio $R$ for $J/\psi + \eta_c + X$ at this order includes two terms:
\begin{eqnarray}
R[J/\psi + \eta_c + X] &=& 
\frac{2 \pi^2 \alpha_s^2}{ 9} \,
X^2 (r^2 - Y)^2 r^2 (1-r^2)^{3/2} \,
\frac{\langle O_1^{J/\psi}(^3S_1) \rangle }{ 3 m_c^3} \,
\frac{\langle O_1^{\eta_c}(^1S_0) \rangle }{ m_c^3} 
\nonumber
\\
&+& \sum_{J=0}^2
\frac{\pi^2 \alpha_s^2 }{ 108} \,
X_8^2 G_J(r) r^4 (1-r^2)^{3/2} \,
\frac{\langle O_8^{J/\psi}(^3P_J) \rangle }{ (2J+1)m_c^5} \,
\frac{\langle O_8^{\eta_c}(^1P_1) \rangle }{ 3m_c^5} .
\label{color-octet}
\end{eqnarray}
By the velocity-scaling rules of NRQCD \cite{BBL}, 
the color-octet term is suppressed by a power of $v^8$, 
but that is partly compensated by an enhancement factor of $1/r^6$.
Omitting the color-octet term and applying the vacuum-saturation approximation 
to the NRQCD matrix elements in the color-singlet term,
we recover the nonrelativistic limit of the ratio $R$ in Eq.~(\ref{R:psi+eta})
for the exclusive $J/\psi + \eta_c$ final state.

Color-octet processes can also contribute to the exclusive cross section for
$J/\psi + \eta_c$.  The 2 gluons emitted by one color-octet $c \bar c$ pair 
in the transition to a color-singlet state that can form charmonium 
can be absorbed by the other color-octet $c \bar c$ pair.
The amplitude is suppressed by $v^4$ relative to the color-singlet amplitude.
The leading contribution to the cross section will come from the interference 
between these two amplitudes and so will be suppressed only by $v^4$.
The interference terms cancel when summed over all possible final states,
so they do not appear in the inclusive cross section (\ref{color-octet}).
The color-octet contributions will also have suppression factors of $r^2$
that guarantee consistency with the helicity selection rules of perturbative 
QCD in the limit $r \to 0$.

\subsection{ Phenomenology
	\label{sec:predict}}

The BELLE Collaboration has recently measured the cross section 
for $J/\psi + \eta_c$ \cite{Abe:2002rb}.
The $J/\psi$ was detected
through its decays into $\mu^+\mu^-$ and $e^+e^-$, 
which have a combined branching fraction of about 12\%.
The $\eta_c$ was observed as a peak in 
the momentum spectrum of the $J/\psi$ corresponding to the 2-body 
process $J/\psi + \eta_c$.  
The measured cross section is
\begin{eqnarray}
\sigma[J/\psi+\eta_c] \times B[\ge 4] 
= \left( 33^{+7}_{-6} \pm 9 \right) \; {\rm fb},
\label{Belle}
\end{eqnarray}
where $B[\ge 4]$ is the branching fraction for the $\eta_c$ to decay 
into at least 4 charged particles.  Since $B[\ge 4]<1$, 
the right side of Eq.~(\ref{Belle}) is a lower bound on the cross section 
for $J/\psi + \eta_c$. 
 
The lower bound provided by Eq.~(\ref{Belle}) is about an order of magnitude 
larger than the central value 3.8 fb of the calculated cross section 
for $J/\psi + \eta_c$ in Table~\ref{tab:sigma}.
The largest theoretical errors are  QCD perturbative corrections, 
which we estimate to give an uncertainty of roughly 60\%, 
the value of $m_c$, 
which we estimate to give an uncertainty of roughly 50\%,
and a relativistic correction that we have not been able to quantify
with confidence.
If we take the calculations of the relativistic corrections in 
Table~\ref{tab:sigma-rel} seriously, the extreme upper end of the 
prediction is marginally compatible with the BELLE measurement.
In our further discussion, we will ignore the large discrepancy between the
predicted cross section for $J/\psi + \eta_c$ and the BELLE measurement.
We will focus on predictions for the ratios of cross sections 
from Table~\ref{tab:sigma} under the assumption that many of the 
theoretical errors will cancel in the ratio.

In addition to measuring the cross section for $J/\psi + \eta_c$,
the BELLE Collaboration saw evidence for $J/\psi + \eta_c(2S)$ and 
$J/\psi + \chi_{c0}(1P)$ \cite{Abe:2002rb} events.  A 3-peak fit to the 
momentum spectrum of the $J/\psi$ gave approximately 67, 42, and 39 events 
for $\eta_c$, $\eta_c(2S)$, and $\chi_{c0}(1P)$ with fluctuations 
of 12-15 events.  The proportion of $J/\psi + \eta_c$, 
$J/\psi + \eta_c(2S)$, and $J/\psi + \chi_{c0}(1P)$ events, 
$1.00 \, : \, 0.63\pm 0.25 \, : \, 0.58\pm 0.24$,
is consistent with the
proportions  $1.00 \, : \, 0.41 \, : \, 0.63$ of the cross sections
in Table~\ref{tab:sigma}.
These proportions are insensitive to the choice of $m_c$.
The absence of 
peaks corresponding to $\chi_{c1}(1P)$ and $\chi_{c2}(1P)$ is also
consistent with Table~\ref{tab:sigma}.  The cross sections for 
$J/\psi + \chi_{cJ}(1P)$ for $J=1$ and 2 are predicted to be smaller 
than for $J=0$ by factors of about 0.16 and 0.29, respectively.

If the cross sections for the narrow $D$-wave states are large enough,
they could be discovered at the $B$ factories.
The state $\eta_{c2}(1D)$ could be observed as a peak in the
momentum spectrum of $J/\psi$ corresponding to the 2-body 
process $J/\psi + \eta_{c2}(1D)$.  
The prediction in Table~\ref{tab:sigma} for the cross section for 
$J/\psi + \eta_{c2}(1D)$ is smaller than that for 
$J/\psi + \eta_c$ by about a factor 0.09. 
It might also be possible to discover the $D$-wave state $\psi_2(1D)$ 
as a peak in the momentum spectrum of $\eta_c$ corresponding to the 2-body 
process $\psi_2(1D) + \eta_c$. The $\eta_c$ could be detected through 
its decay into $K K \pi$, whose branching fraction is about 6\%.
The prediction in Table~\ref{tab:sigma} for the cross section for 
$\psi_2(1D) + \eta_c$ is smaller than that for 
$J/\psi + \eta_c$ only by about a factor 0.27. 
Our predictions for the cross sections for $J/\psi + \eta_{c2}(1D)$ and
$\psi_2(1D) + \eta_c$ are based on a phenomenological determination 
of the NRQCD matrix elements that ignored mixing between the
$\psi(2S)$ and the $\psi_1(1D)$.  If there is significant mixing between
these two states, the cross sections for $S$-wave + $D$-wave
could be a factor of 7 smaller.

In summary, we have calculated the cross sections for 
$e^+ e^-$ annihilation into exclusive double charmonium states
with opposite charge conjugation.
Many of the cross sections are large enough to be observed at $B$ factories.
In particular, it may be possible to discover the $D$-wave states
$\eta_{c2}(1D)$ and $\psi_2(1D)$.  The cross sections for double charmonium
suffer from fewer theoretical uncertainties than inclusive
charmonium cross sections.
The largest uncertainty comes from relativistic corrections.
Measurements of exclusive double charmonium cross sections will provide
strong motivation for developing reliable methods for calculating the
relativistic corrections to quarkonium cross sections.

{\bf Note added:}
Liu, He, and Chao have calculated the $\alpha^2 \alpha_s^2$ terms
in the cross sections for $e^+ e^-$ annihilation into 
$J/\psi + H$, $H = \eta_c, \chi_{c0}, \chi_{c1}, \chi_{c2}$ \cite{Liu:2002wq}.
Their results are consistent with ours.
In Ref.~\cite{Liu:2004ga}, Liu, He, and Chao calculated
the $\alpha^2 \alpha_s^2$ terms in the differential cross sections for 
$J/\psi + \eta_c$, $J/\psi + \chi_{c0}$, and $\eta_c +h_c$.
Their results for $J/\psi + \eta_c$ and $J/\psi + \chi_{c0}$ agree with ours.
Our corrected result for $\eta_c +h_c$ in Eq.~(\ref{R:etac-hc}) 
agrees with their result.

\appendix
\section{Angular distributions}
In this appendix, we give the angular distributions $dR/dx$ for
$e^+ e^- \to H_1(\lambda_1) + H_2(\lambda_2)$
for each of the helicity states that contribute at order $\alpha^2 \alpha_s^2$
or $\alpha^4$.
The angular variable is $x = \cos \theta$, where $\theta$ is the angle 
between $e^-$ and $H_1$ in the $e^+e^-$ center-of-mass frame.
The results for $R$ in the text are obtained by summing over all the 
helicities and integrating over $-1 < x < +1$.

\subsection{$\bm{S}$-wave + $\bm{S}$-wave}
The angular distribution for $J/\psi + \eta_c$ is
\begin{eqnarray}
\frac{dR}{dx}[J/\psi(\pm 1) + \eta_c]
=
\frac{\pi^2 \alpha_s^2 }{24} \,
X^2 (r^2-Y)^2 r^2 (1-r^2)^{3/2} \,
\frac{\langle O_1 \rangle_{J/\psi} \langle O_1 \rangle_{\eta_c} }{ m_c^6}
\,(1+x^2),
\end{eqnarray}
The cross section for the longitudinal helicity component $\lambda_1=0$ 
of the $J/\psi$ vanishes.

\subsection{$\bm{S}$-wave + $\bm{P}$-wave}
The angular distributions for $J/\psi + \chi_{cJ}$ are
\begin{eqnarray}
\frac{dR}{dx}[J/\psi(\lambda_1) + \chi_{cJ}(\lambda_2)]
=
\frac{ \pi^2 \alpha_s^2 }{ 432} \,
X^2  r^2 (1-r^2)^{1/2} \,
\frac{\langle O_1 \rangle_{J/\psi} \langle O_1 \rangle_{\chi_{cJ}} }{ m_c^8} 
F_J(\lambda_1,\lambda_2,x).
\end{eqnarray}
The non-vanishing entries of $F_J(\lambda_1,\lambda_2,x)$ are 
\begin{subequations}
\begin{eqnarray}
F_0(0,0,x)&=&\frac{3}{4}r^2[ 4 + 2(Y+5) r^2 -3 r^4 ]^2 (1-x^2),
\\
F_0(\pm 1,0,x)&=& \frac{3}{8}[ 4 Y -6 (Y+3) r^2 +7 r^4 ]^2(1+x^2),
\\
F_1(0,\pm 1,x)&=&\frac{9}{16} r^4 [ 2(3 Y + 4) -7 r^2 ]^2 (1+x^2),
\\
F_1(\pm 1,0,x)&=&\frac{9}{16}[ 8 Y - 2Y r^2 + r^4 ]^2 (1+x^2),
\\
F_1(\pm 1,\pm 1,x)&=&\frac{9}{8} r^2 [ 4 Y + 2(Y+2) r^2 - 3 r^4 ]^2 (1-x^2),
\\
F_2(0,0,x)&=&\frac{3}{2}r^2 [ 4   + 2 (Y -1) r^2 - 3 r^4 ]^2(1-x^2),
\\
F_2(0,\pm 1,x)&=&\frac{9}{16}r^4 [ 2(Y+2) - 5r^2 ]^2 (1+x^2),
\\
F_2(\pm 1,0,x)&=&\frac{3}{16} [8 Y - 6 (Y+2) r^2 + 11 r^4  ]^2(1+x^2),
\\
F_2(\pm 1,\pm 1,x)&=&\frac{9}{8}r^2 [ 4 Y - 2 (Y +2) r^2 + 3 r^4 ]^2(1-x^2),
\\
F_2(\pm 1,\pm 2,x)&=&\frac{9}{8}r^4 [ 2Y-r^2 ]^2(1+x^2).
\end{eqnarray}
\end{subequations}

The angular distribution for $\eta_c+h_c$ is
\begin{equation}
\frac{dR}{dx}\left[
\eta_c+h_c(\lambda)
\right]
=
 \frac{\pi^2 \alpha_s^2} {144} X^2 r^4
(1-r^2)^{1/2} \frac{\langle O_1 \rangle_{\eta_c}
                    \langle O_1 \rangle_{h_c} }{m_c^8} H(\lambda,x) ~,
\end{equation}
where
\begin{subequations}
\begin{eqnarray}
H(0,x)&=&\frac{3}{4}(1-x^2)(3r^4-6r^2+4)^2,
\\
H(\pm 1,x)&=&\frac{3}{8}(1+x^2)r^2(r^2-2)^2.
\end{eqnarray}
\end{subequations}

\subsection{$\bm{P}$-wave + $\bm{P}$-wave}
The angular distributions for $h_c + \chi_{cJ}$ are
\begin{eqnarray}
\frac{dR}{dx}[h_c(\lambda_1)+\chi_{cJ}(\lambda_2)]
=
\frac{\pi^2 \alpha_s^2            }{ 108} \,
X^2 r^4 (1-r^2)^{3/2} \,
\frac{\langle O_1 \rangle_{h_c} \langle O_1 \rangle_{\chi_{cJ}} }{ m_c^{10}}
G_J(\lambda_1,\lambda_2,x).
\end{eqnarray}
where non-vanishing entries of $G_J(\lambda_1,\lambda_2,x)$
are given by
\begin{subequations}
\begin{eqnarray}
G_0(\pm 1,0,x)
&=&
\frac{3}{8}r^2(6-r^2)^2(1+x^2),
\\
G_1(0,0,x)&=&18(1-x^2),
\\
G_1(0,\pm 1,x)&=&\frac{225}{16}r^2(2-r^2)^2(1+x^2),
\\
G_1(\pm 1,0,x)&=&\frac{9}{16}r^2(2-r^2)^2(1+x^2),
\\
G_2(0,\pm 1,x)&=&\frac{9}{16}r^2(4-5r^2)^2(1+x^2),
\\
G_2(\pm 1,0,x)&=&\frac{3}{16}r^6(1+x^2),
\\
G_2(\pm 1,\pm 2,x)&=&\frac{9}{8}r^6(1+x^2).
\end{eqnarray}
\end{subequations}
Note that $G_2(\pm 1,0,x)$ is more suppressed than the prediction from 
the hadron helicity conservation rule.
\subsection{$\bm{S}$-wave + $\bm{D}$-wave}
The angular distributions for $J/\psi + \eta_{c2}$ are
\begin{eqnarray}
\frac{dR}{dx}
[J/\psi(\pm 1)+ \eta_{c2}(0)] =
\frac{\pi^2 \alpha_s^2           }{36} \,
X^2 (Y-2r^2)^2 r^2 (1-r^2)^{7/2} \,
\frac{\langle O_1 \rangle_{J/\psi} \langle O_1 \rangle_{\eta_{c2}} }{ m_c^{10}}
(1+x^2).
\end{eqnarray}
The cross sections for the longitudinal helicity component $\lambda_1=0$ 
of the $J/\psi$ and the helicity components 
$\lambda_2=\pm 1$ and $\pm 2$ of the $\eta_{c2}$ vanish.

The angular distributions for $\psi_1 + \eta_c$ are
\begin{eqnarray}
\frac{dR}{dx}[\psi_1(\pm 1)+\eta_c] =
\frac{\pi^2 \alpha_s^2            }{23040} \,
X^2 (26Y-21r^2+10r^4)^2 r^2 (1-r^2)^{3/2} \,
\frac{\langle O_1 \rangle_{\psi_1} \langle O_1 \rangle_{\eta_c} }%
     { m_c^{10}}(1+x^2).
\label{dRdx:3D1+eta}
\end{eqnarray}
The cross section for the longitudinal helicity component $\lambda_1=0$ 
of the $\psi_1$ vanishes.

The angular distributions for $\psi_2 + \eta_c$ are
\begin{eqnarray}
\frac{dR}{dx}
[\psi_2(\lambda_1)+\eta_c] =
\frac{\pi^2 \alpha_s^2            }{ 54} \,
X^2 r^4 (1-r^2)^{3/2} \,
\frac{\langle O_1 \rangle_{\psi_2} \langle O_1 \rangle_{\eta_c} }{ m_c^{10}}
H(\lambda_1,x).
\end{eqnarray}
The non-vanishing entries of $H(\lambda_1,x)$ are 
\begin{subequations}
\begin{eqnarray}
H(0,x)
&=& 
\frac{9}{2}(1-x^2),
\\
H(\pm 1,x)
&=&
\frac{3}{16} r^2(7-4r^2)^2 (1+x^2).
\end{eqnarray}
\end{subequations}

\begin{acknowledgments}
We thank G.~T.~Bodwin and E.~Eichten for many useful discussions.
We acknowledge B.~Yabsley for a suggestion
that led to our adding the Appendix. 
We thank Chaehyun Yu for his help in checking some of the calculations.
The research of E.B.~is supported in part by the U.~S.~Department of Energy,
Division of High Energy Physics, under grant DE-FG02-91-ER4069
and by Fermilab, which is operated by Universities Research Association Inc.
under Contract DE-AC02-76CH03000 with the U.~S.~Department of Energy.
The research of J.L.~in the High Energy Physics Division
at Argonne National Laboratory is supported by
the U.~S.~Department of Energy, Division of High Energy Physics, under
Contract W-31-109-ENG-38.
\end{acknowledgments}


\end{document}